\documentclass[prb, twocolumn, showpacs]{revtex4}
\usepackage{epsfig}
\usepackage[none]{hyphenat}
\newcommand{\ben}{\begin{eqnarray}}
\newcommand{\een}{\end{eqnarray}}

\newcommand{\bef}{\begin{figure}[h!bt]\centering}
\newcommand{\eef}{\end{figure}}
\newcommand{\bet}{\begin{table}[hbt]\centering}
\newcommand{\eet}{\end{table}}

\begin{document}
%%%%%%%%%%%%%%%%%%%%%%%%%%%%%%%
\title{Temperature-concentration phase diagram of (Ca$_{1-x}$La$_x$)$_{10}$(Pt$_3$As$_8$)(Fe$_2$As$_2$)$_5$ superconductors}
\author{N. Ni$^{1,2}$, W. E. Straszheim$^3$, D. J. Williams$^1$, M. A. Tanatar$^3$, R. Prozorov$^3$, E. D. Bauer$^1$, F. Ronning$^1$, J. D. Thompson$^1$ and R. J. Cava$^2$}
\affiliation{
$^1$Los Alamos National Laboratory, Los Alamos, NM 87544, USA\\
$^2$Department of Chemistry, Princeton University, Princeton, NJ 08544, USA\\
$^3$Ames Laboratory and Department of Physics and Astronomy, Iowa State University, Ames, Iowa 50011, USA\\
}

\begin{abstract}
Single crystals of (Ca$_{1-x}$La$_x$)$_{10}$(Pt$_3$As$_8$)(Fe$_2$As$_2$)$_5$ ($x=0$ to 0.182) superconductors have been grown and characterized by X-ray, microprobe, transport and thermodynamic measurements. Features in the magnetic susceptibility, specific heat and two kinks in the derivative of the electrical resistivity around 100 K in the $x=0$ compound support the existence of decoupled structural and magnetic phase transitions. With La doping, the structural/magnetic phase transitions are suppressed and a half-dome of superconductivity with a maximal $T_c$ around 26 K is observed in the temperature-concentration phase diagram.

\end{abstract}
\date{\today}
\maketitle
%%%%%%%%%%%%%%%%%%%%%%%%%%%%%%%
The report of superconductivity at 26 K in LaFeAsO$_{0.9}$F$_{0.1}$ \cite{jacs} has led to the discovery of several families of high $T_c$ iron arsenide superconductors, including the so-called 1111, 122, 111 and 42622 families \cite{rotter, 111, 42622b}.
The intense study of these families has enriched our understanding of the interplay among structure, magnetism and superconductivity. Recently a new iron-arsenide compound, Ca$_{10}$(Pt$_3$As$_8$)(Fe$_2$As$_2$)$_5$ (the so called 10-3-8 compound), has been characterized \cite{pnas, japan, german}. This compound crystallizes in a triclinic structure with space group P -1 and has -Ca-(Pt$_3$As$_8$)-Ca-(Fe$_2$As$_2$)- layer stacking, as shown in the left inset of Fig. \ref{xray}. The FeAs layer is made of edge-sharing FeAs$_4$ tetrahedra, the key structural element in all the Fe-pnictide superconductors. A structural phase transition around 100 K has been revealed in this compound using polarized light imaging \cite{ruslan}. Although the susceptibility drop observed to accompany the long range antiferromagnetic ordering in the 1111 and 122 families has not been reported for the 10-3-8 parent compound \cite{pnas, chen}, recent NMR measurements show that this compound orders antiferromagnetically (AFM) below $\sim$100 K \cite{stuart}. With Pt substitution on the Fe sites, superconductivity up to 12 K has been realized \cite{pnas, chen}. In the 1111 and 122 families, doping on the intermediary layer results in a higher T$_c$ than doping on the FeAs layers, and thus higher T$_c$ may be expected for intermediary layer doping in the 10-3-8 compound as well. Indeed, 20\% La doping on the Ca sites in this compound was found to show a $T_c$ of 30 K \cite{dirk}. In this paper, we report the systematic characterization of (Ca$_{1-x}$La$_x$)$_{10}$(Pt$_3$As$_8$)(Fe$_2$As$_2$)$_5$ single crystals via X-ray diffraction, microprobe, transport and thermodynamic measurements. Due to improved quality of the single crystals, we are able to observe a resistivity jump, a susceptibility drop, and a specific heat jump in the parent 10-3-8 compound, supporting the existence of both structural and magnetic phase transitions. A $T-x$ phase diagram for (Ca$_{1-x}$La$_x$)$_{10}$(Pt$_3$As$_8$)(Fe$_2$As$_2$)$_5$ is presented.
%%%%%%%%%%%%%%%%%%%%%%%%%%%%%%%
\bef \psfig{file=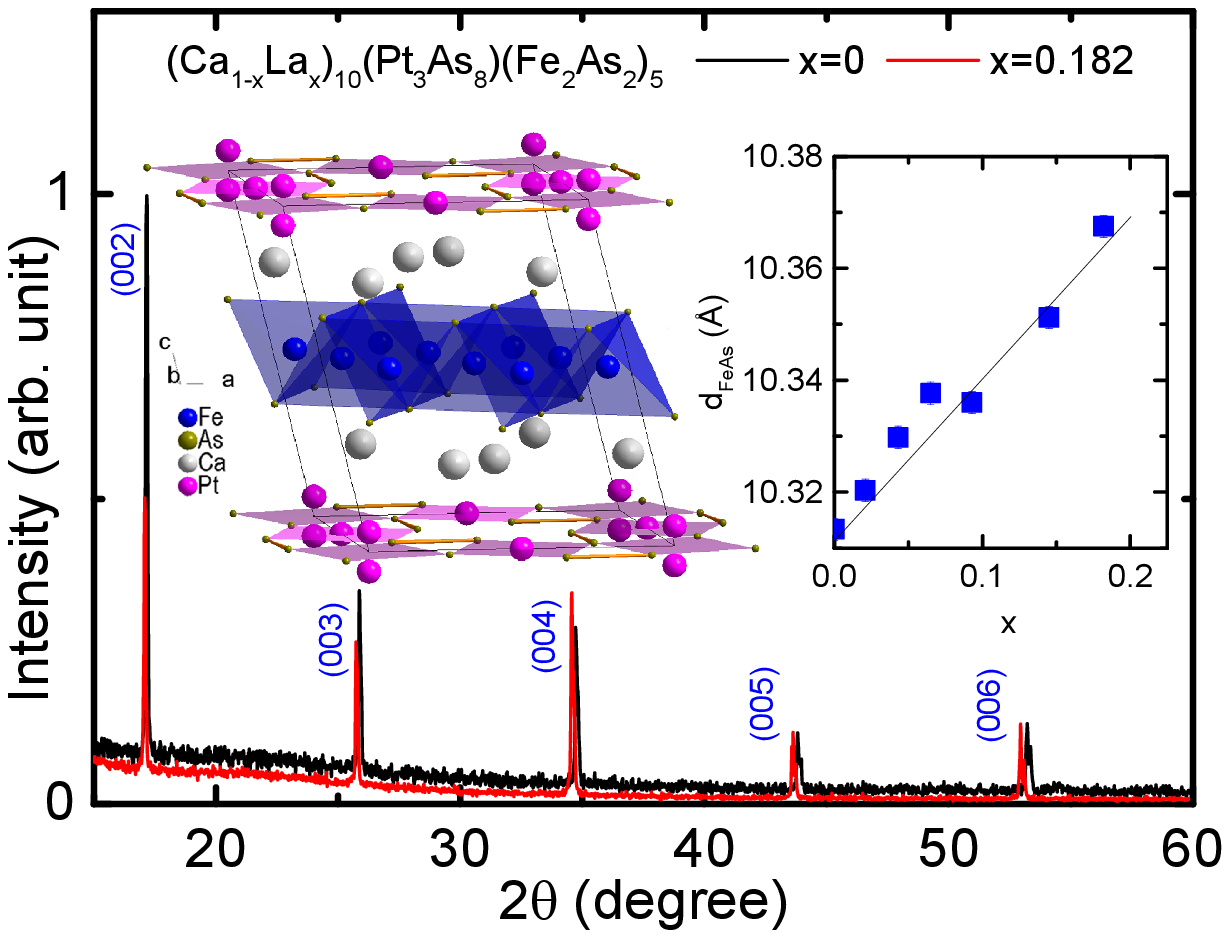,width=3.4in}
\caption{The $(00l)$ diffraction pattern of (Ca$_{1-x}$La$_{x}$)$_{10}$(Pt$_3$As$_8$)(Fe$_2$As$_2$)$_5$ ($x=0, 0.182$). Left inset: The crystal structure of Ca$_{10}$(Pt$_3$As$_8$)(Fe$_2$As$_2$)$_5$. Right inset: The interlayer FeAs distance vs. doping level $x$.} \label{xray} \eef
%%%%%%%%%%%%%%%%%%%%%%%%%%%%%%%

 \begin{table}
\caption{The nominal Ca:La:Fe:Pt:As ratio in the crystal growth. WDS measured doping level $x$.} \label{tab:data}
\medskip
\centering
\begin{tabular}{c|c|c|c|c|c|c|c}
\hline
\hline

   \multicolumn{8}{c}{(Ca$_{1-x}$La$_x$)$_{10}$(Pt$_3$As$_8$)(Fe$_2$As$_2$)$_5$}\\
     \hline
     \hline

       nominal Ca & 3.5 & 3.45& 3.4 & 3.35 & 3.3 & 4.1 & 4.2 \\
        nominal La  &0   &0.05 &0.1  &0.15  &0.2  &0.4  &0.8\\
        nominal Fe &2    &2    &2    &2     &2    &2    &2\\
        nominal Pt &0.4  &0.4  &0.4  &0.4   &0.4   &0.4 &0.4\\
        nominal As &5.5  &5.5  &5.5  &5.5   &5.5   &6.5  &7\\

   $x$ & 0 & 0.021 &0.043  & 0.065 & 0.093 & 0.145&0.182 \\

\hline

\hline
\end{tabular}
\end{table}

Plate-like millimeter-sized single crystals were successfully grown from a CaAs-rich flux\cite{pnas}. CaAs, FeAs and LaAs precursors were made using the solid state reaction method. These precursors and Pt powder were mixed thoroughly according to the nominal ratios listed in Table I. The mixture was pressed into a pellet, put in an Al$_2$O$_3$ crucible and sealed into a quartz tube under vacuum. The resulting ampules were heated up to $1150^o C$, held for 96 hours, slowly cooled down to $885^o C$ and then quenched. After rinsing off the flux using distilled water, single crystals were obtained. These single crystals show a layered growth habit and are easily exfoliated and bent. In each batch, small crystals, with thickness less than 0.03 mm, were employed in the transport measurements. The La concentration $x$ was obtained via wavelength dispersive spectroscopy (WDS) using the electron probe microanalyzer of a JEOL JXA-8200 electron-microprobe. WDS was performed on the measured transport samples to provide a reliable determination of the electronic phase diagram as a function of composition. The results of the WDS measurements are summarized in Table I. These measurements directly indicate that La has been successfully doped on the Ca sites, while the Pt substitution on the FeAs layer is well controlled: it is 0 for the $x$=0 compound, 0.007 for the $x$=0.021, 0.043, 0.065, and 0.145 compounds and 0.02 for the $x$=0.093 and 0.182 compounds. In this paper, $x$ refers to the WDS value. Transport and specific heat measurements were performed in a Quantum Design (QD) Physical Properties Measurement System. Magnetic properties were measured in a QD Magnetic Properties Measurement System. To easily compare the physical properties of these superconductors with other iron arsenide superconductors, the units of molar susceptibility, magnetization, and heat capacity presented are normalized to per mole-Fe$_2$.

 X-ray diffraction was performed on a Scintag X$_1$ Advances Diffraction System employing Cu $K_{\alpha}$ ($\lambda = 1.5406 \AA\ $) radiation. No FeAs, PtAs$_2$ or other impurities were observed in any X-ray pattern. Figure \ref{xray} shows the $(00L)$ diffraction pattern of (Ca$_{1-x}$La$_{x}$)$_{10}$(Pt$_3$As$_8$)(Fe$_2$As$_2$)$_5$ ($x=0, 0.182$). A peak shift between these two samples is observed. By refining the (00L) diffraction patterns via the UnitCell software \cite{unitcell} the interlayer distance of the FeAs layers were obtained. The right inset of Fig. \ref{xray} shows the evolution of the interlayer FeAs distance with $x$. This distance increases monotonically with La doping from 10.313(2) $\AA$ in the parent compound to 10.368(2) $\AA$ in the $x$=0.182 compound, providing further evidence that La is incorporated into the structure.
%%%%%%%%%%%%%%%%%%%%%%%%%%%%%%%
\bef \psfig{file=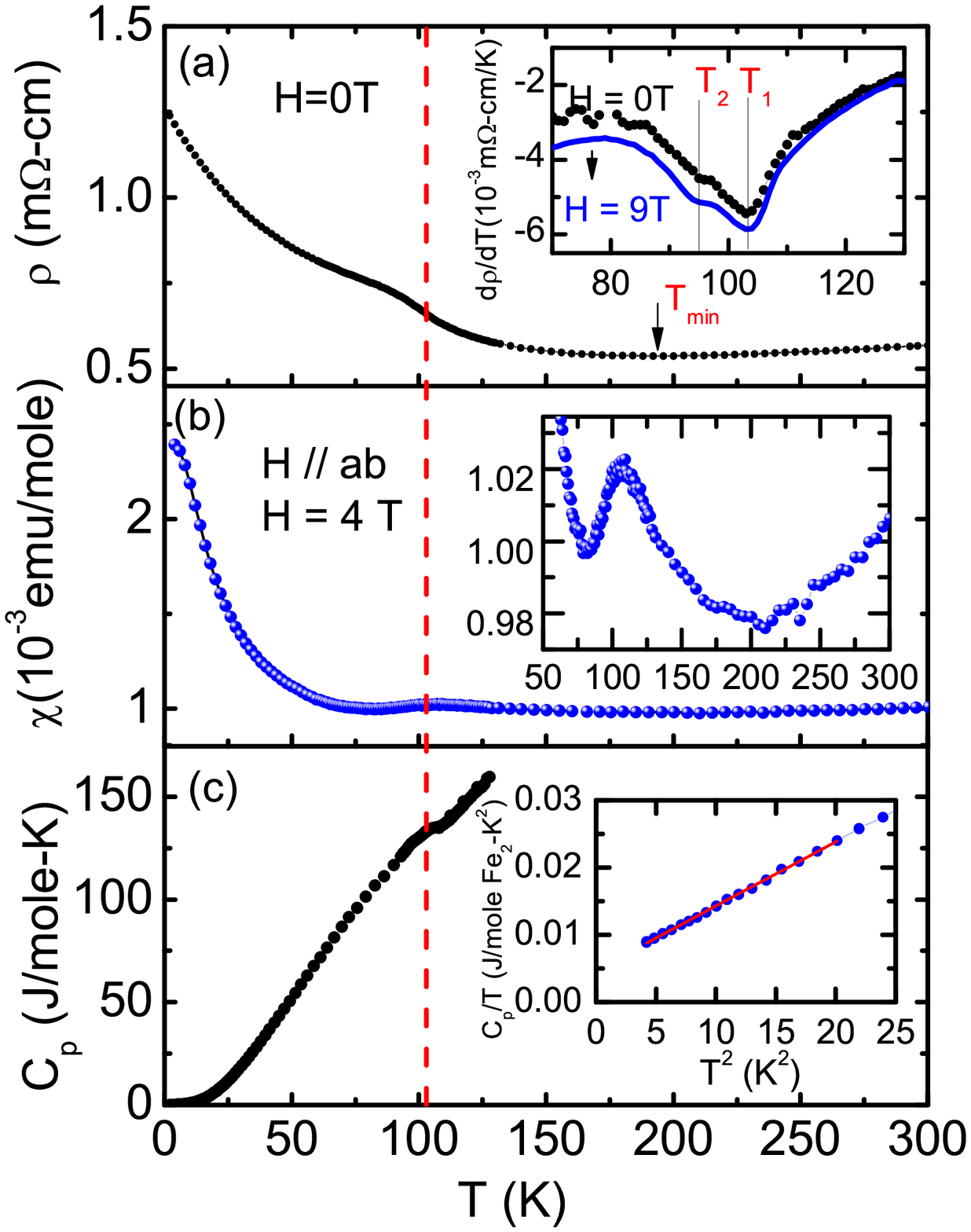,width=3.4in}
\caption{The physical properties of Ca$_{10}$(Pt$_3$As$_8$)(Fe$_2$As$_2$)$_5$. (a) Electrical resistivity $\rho$ (T) taken at H=0T. Inset: d$\rho$/d$T$ vs. $T$ taken at 0 T and 9 T; $T_1$=103 K and $T_2$=95 K. (b) $\chi$ (T) taken at 4 T with $H//ab$. Inset: The $\chi$(T) from 50 to 300 K. (c) $C_p$ vs. $T$. Inset: $C_p/T$ vs. $T^2$ } \label{parent} \eef
%%%%%%%%%%%%%%%%%%%%%%%%%%%%%%%

The physical properties of the parent compound are summarized in Fig. \ref{parent}. Figure \ref{parent} (a) shows the temperature dependence of the resistivity. The resistivity is 0.57 $m\Omega-cm$ at 300 K, which is almost twice as that of BaFe$_2$As$_2$ \cite{makariy1, makariy2}. $\rho$(T) shows a resistivity minimum at $T_{min}$. $T_{min}$ is sample dependent, varying from larger than 300 K to 170 K, with an average of 210 K. It is unclear wether $T_{min}$ comes from disorder or other mechanisms, such as charge gap formation. With decreasing temperature, an abrupt resistivity increase with a bump feature occurs below $\sim 100$ K. No hysteresis is observed between zero field cooling and warming $\rho$(T) data. The inset shows the temperature derivative of the electrical resistivity $d\rho/dT$ obtained at $\rm H=0$ and 9 T. The parent 10-3-8 compound shows two kinks at $T_1$ = 103 K and $T_2$ = 95 K in $d\rho/dT$. These features are also observed in underdoped Ba(Fe$_{1-x}$Co$_x$)$_2$As$_2$ \cite{niall}, where the higher temperature kink is related to the structural phase transition
and the lower temperature kink is related to the magnetic
phase transition  \cite{dan}, suggesting that a structural phase transition in the parent 10-3-8 phase may occur at 103 K and a magnetic phase transition may occur at 95 K. No change in the temperature of the anomalies is observed with 9 T applied field. Figure \ref{parent} (b) shows the temperature dependent susceptibility $\chi$ (T) taken at 4 T. At 300 K, the susceptibility is around 1$\times 10^{-3}$emu/mole, similar to that of

BaFe$_2$As$_2$. Unlike the linear temperature dependence observed in BaFe$_2$As$_2$ \cite{linear, chenlinear} at high temperatures, from 300 to 100 K, the susceptibility is only weakly temperature dependent with a minimum around 200 K. As temperature decreases, a drop in susceptibility is observed at $\sim 100$ K, which is shown in the inset of Fig. \ref{parent} (b). This susceptibility drop is consistent with the resistivity measurement and supports the existence of a magnetic/structural phase transition despite the highly two-dimensional crystal structure. Below 80 K, a Curie tail is observed, which may be caused by magnetic impurities. This paramagnetic contribution combines with the intrinsic magnetism and may lead to the weakly temperature dependent susceptibility from 200 to 300 K. Figure \ref{parent} (c) shows the temperature dependent specific heat data. A clear specific heat anomaly is observed around 100 K, which is consistent with both $\rho$(T) and $\chi$(T) data. Below 5 K, assuming there are no magnetic excitations, $C_p/T$ obeys the relation of $C_p/T$=$\gamma$ +$\beta$ T$^2$, with an electronic specific heat coefficient $\gamma=4.8$ mJ/mole-Fe$_2$-K$^2$ and $\beta=0.95$ mJ/mole-Fe$_2$-K$^3$, corresponding to a Debye temperature of 256 K.
%%%%%%%%%%%%%%%%%%%%%%%%%%%%%%%
\bef \psfig{file=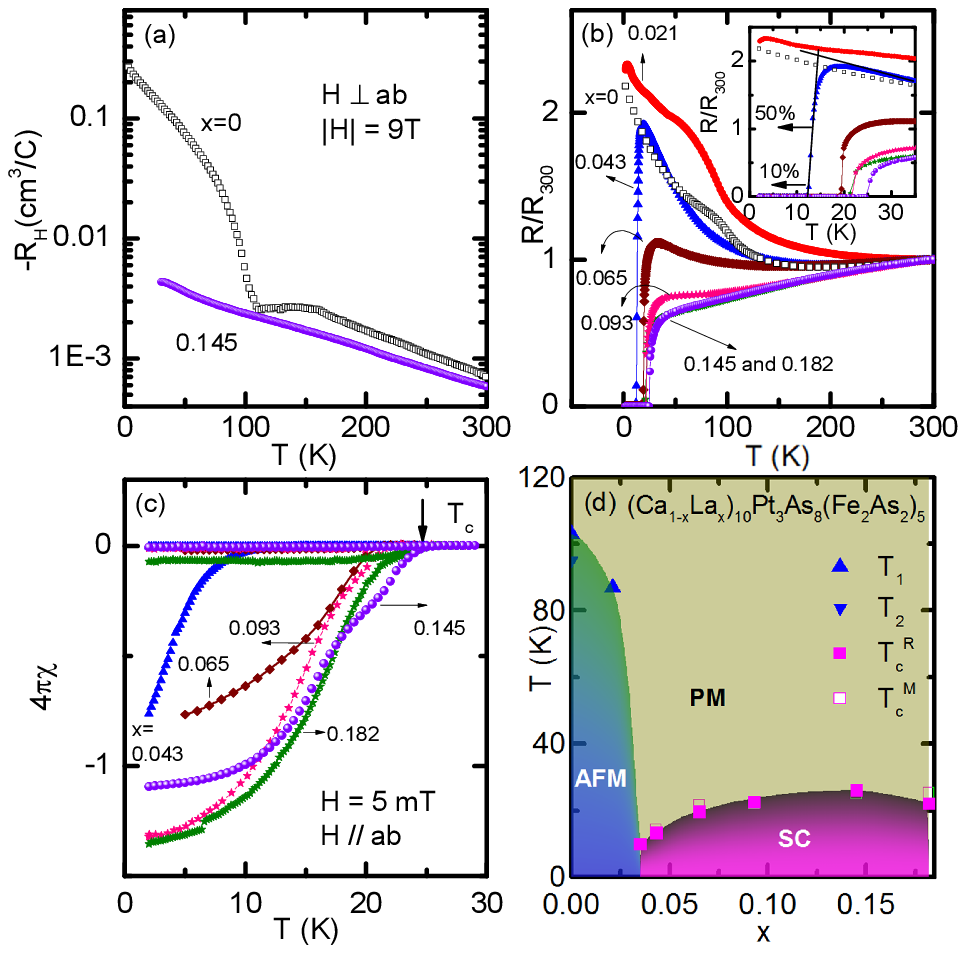,width=3.5in}
\caption{The evolution of the (Ca$_{1-x}$La$_x$)$_{10}$(Pt$_3$As$_8$)(Fe$_2$As$_2$)$_5$ series with doping. (a) negative Hall coefficient -$R_H$(T) for $x=0$ and 0.145 compositions. (b) The temperature dependent $R/R_{300K}$. Inset: The amplified $R/R_{300K}$. (c) The ZFC and FC 4$\pi\chi$ taken at 5 mT with H along the $ab$ plane. (d) The temperature-concentration phase diagram.
} \label{trend} \eef
%%%%%%%%%%%%%%%%%%%%%%%%%%%%%%%

The evolution of the (Ca$_{1-x}$La$_x$)$_{10}$(Pt$_3$As$_8$)(Fe$_2$As$_2$)$_5$ series with doping is presented in Fig. \ref{trend}.
Figure \ref{trend} (a) shows the temperature dependent Hall coefficient for the $x$=0 and $x$=0.145 compounds. The negative Hall coefficient indicates the dominant role of electrons. A dramatic slope change of $Log_{10}|R_H|$, indicating a gap opening related to the structural/magnetic phase transition, is observed in the parent compound, but not in the $x$=0.145 compound. Within the single band model, the carrier concentration is determined from $n$ = -1/$eR_H$, which leads to
$n_{300K}^{x=0}$=8.7$\times 10^{21}$ $\rm cm^{-3}$ and $n_{300K}^{x=0.145}$=1.06$\times 10^{22}$$\rm cm^{-3}$. Using the unit cell volume V = 788.1 ${\AA}^3$ \cite{pnas}, the estimated extra carrier concentration due to the La doping is ($n_{300K}^{x=0.145}$-$n_{300K}^{x=0}$)$\times$V = 1.5/unit cell. This number is consistent with the WDS measurement assuming one La atom adds one electron. Figure \ref{trend} (b) shows the temperature dependent normalized resistivity $R/R_{300 K}$. The high temperature resistive bump is suppressed to 87 K in the $x=0.021$ compound. No resistive bump is detected when $x \geq 0.043$. Comparing with the Pt doped 10-3-8 series \cite {pnas, ruslan}, this implies that no structural phase transition occurs. For the samples with $x \leq 0.093$, the normal state resistance shows a resistivity minimum at $T_{min}$. Although $T_{min}$ is sample dependent, its average value decreases with increasing doping and disappears at $x=0.145$. From the pieces we measured, the average $T_{min}$ is: 210$\pm 30$ K for $x=0$, 200$\pm 15$ K for $x=0.021$, 180$\pm 30$ K for $x=0.043$, 130$\pm 40$ K for $x=0.065$ and 70$\pm 5$ K for $x=0.093$. Superconductivity occurs when $x\geq0.043$. Using the 50\% criterion shown in the inset of Fig. \ref{trend} (b), $T_c$ first appears at 13.3 K in the $x=0.043$ sample, rises to 26.1 K in the $x=0.145$ sample, and then decreases to 22.1 K in the $x=0.182$ sample. Figure \ref{trend} (c) presents ZFC and FC susceptibility data taken at 5 mT with $H//ab$. The criterion to infer $T_c$ is shown in the figure. Although the transitions are broader compared to Ba(Fe$_{1-x}$Co$_x$)$_2$As$_2$ single crystals, the large shielding fraction is comparable to the 122 series, which indicates bulk superconductivity. A small Meissner fraction is a common feature in Fe-pnictide superconductors and is attributed to flux pinning. The temperature-composition phase diagram, constructed from the above physical properties, is shown in Fig. \ref{trend} (d). With La doping, the structural/magnetic phase transitions are suppressed and superconductivity occurs, with a maximum $T_c$ around 26.1 K at $x=0.145$. Due to difficulty in controlling the doping level precisely around $x=0.03$, it is not yet clear whether there is a coexistence region of AFM and superconductivity, nor is it clear how $T_2$ evolves with doping. Further work is necessary to resolve these issues.

%%%%%%%%%%%%%%%%%%%%%%%%%%%%%%%
\bef \psfig{file=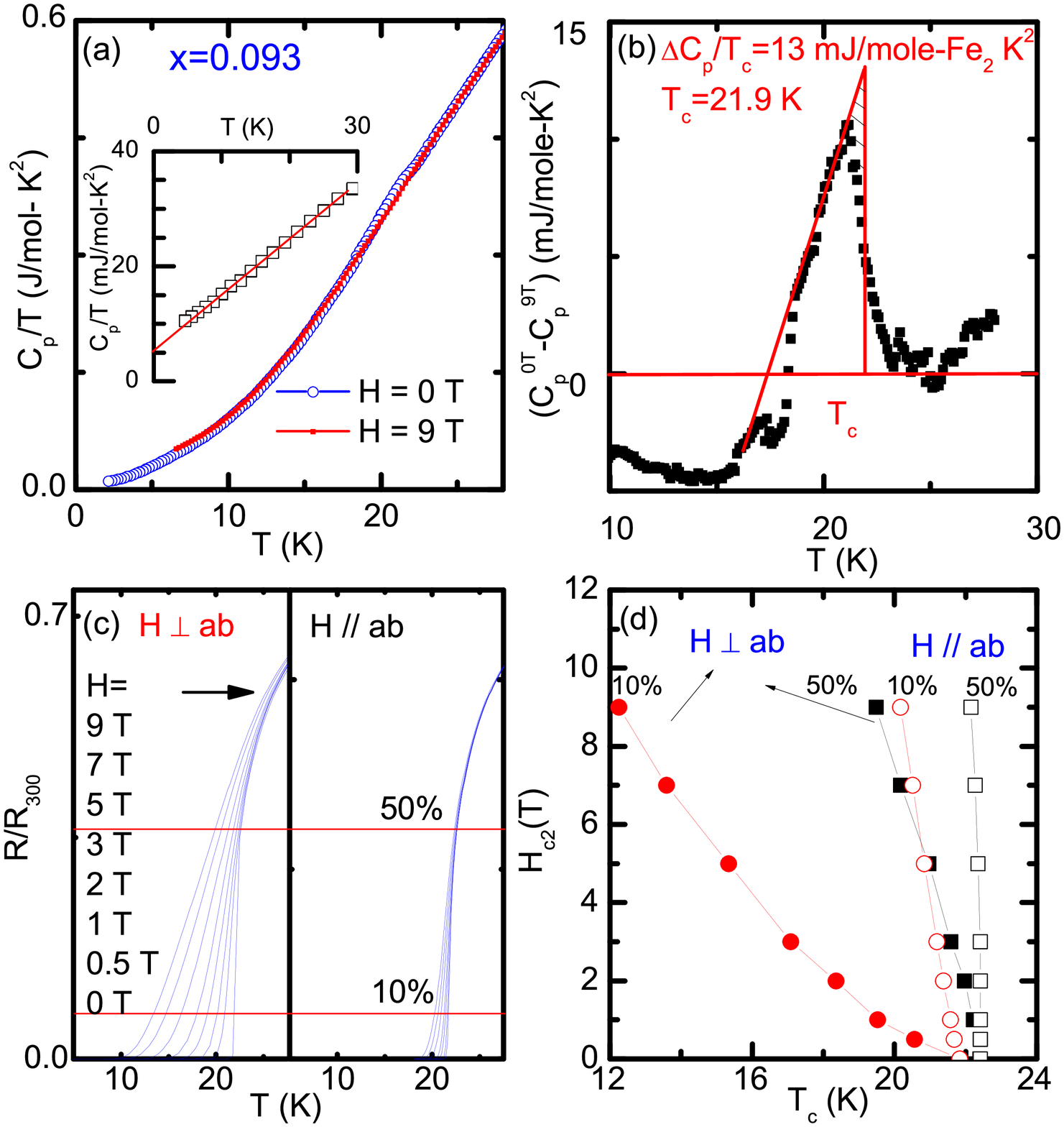,width=3.4in}
\caption{Physical properties of the superconducting state in the $x$=0.093 10-3-8 compound. (a) $C_p/T$ vs. T in $H=0$ T (open squares) and 9 T (solid line). Inset: $C_p/T$ vs. $T^2$ in 0 T. (b) $(C_p^{0T}-C_p^{9T})/T$ vs. T. $T_c$ is inferred using the equal entropy construction shown in figure (c) $R/R_{300K}$ taken at H=0, 0.5, 1, 2, 3, 5, 7, 9 T with H along and perpendicular to the $ab$ plane. (d) $H_{c2}$ vs. T obtained with the 10\% and 50\% criteria from the resistivity data in (c).} \label{sc} \eef
%%%%%%%%%%%%%%%%%%%%%%%%%%%%%%%

To study the superconducting state of the La doped 10-3-8 superconductors in detail, a representative sample with $x=0.093$ was chosen. Figure \ref{sc} (a) shows the specific heat measured at $\rm H=0$ T and 9 T with $H\bot ab$. A specific heat jump can be observed, confirming bulk superconductivity. The inset shows $C_p/T$ vs. $T^2$ taken at $\rm H= 0$ T. The inferred residual $\gamma$ is 5.8 mJ/mole-$\rm Fe_2$ $K^2$ and the Debye temperature is 257 K. Figure \ref{sc} (b) shows a plot of $(C_p^{0T}-C_p^{9T})/T$ vs. T. Using an equal entropy construction shown in Fig. \ref{sc} (b), the resulting $\Delta C_p/T_c$ is 13 mJ/mole-$\rm Fe_2$ $\rm K^2$ and $T_c$ is 21.9 K. These values fall onto the Bud'ko-Ni-Canfield (BNC) log-log plot reasonably well \cite{sergey, stewart, jkim}, adding one more example to the BNC scaling, which reveals $\Delta C_p/T_c$ is proportional to $T_c^2$ for a large number of 122, 111, 1111 based superconductors. The calculated $\gamma_n$ \cite{gamma} is 16$\pm$2 mJ/mole-$\rm Fe_2$ $\rm K^2$, leading to $\Delta C_p/T_c \gamma_n$ $\sim 0.8$. Figure \ref{sc} (c) shows the suppression of $T_c$ under an applied magnetic field, in which the resistive transition becomes much broader, indicating strong thermal fluctuations of vortices in this compound. Since $T_c$ is suppressed by less than 0.1 K using the 90\% criterion when 9 T is applied along $ab$ plane, only the 50\% and 10\% criteria are employed to infer $T_c$ under field. The derived upper critical field $H_{c2}$(T) is summarized in Fig. \ref{sc} (d). The orbital limiting $H_{c2}(0)$ can be calculated via the WHH equation, -$0.69T_cdH_{c2}/dT|_{T_c}$. Using the 50\% criterion, the estimated $H_{c2}^{//ab}$(0) $\sim$ 400 T and $H_{c2}^{\bot ab}$(0) $\sim$ 60 T; using the 10\% criterion, the estimated $H_{c2}^{//ab}$(0) $\sim$ 70 T and $H_{c2}^{\bot ab}$(0) $\sim$ 10 T. $H_{c2}^{//ab}$(0) obtained from 50\% criterion is almost 4 times of that in SmFeAsO$_{0.8}$F$_{0.2}$ with a $T_c$ of 40 K, implying a very large gap formation. Although the $H_{c2}^{\bot ab}$ curve inferred from the 50\% criterion shows roughly linear behavior, the $H_{c2}^{\bot ab}$ curve inferred from the 10\% criterion shows upward curvature,  which is common in cuprate and multigap 1111 superconductors \cite{cuprate, naturehc2}.

In conclusion, we have characterized superconducting (Ca$_{1-x}$La$_x$)$_{10}$(Pt$_3$As$_8$)(Fe$_2$As$_2$)$_5$ ($x=0$ to 0.182) single crystals. With La doping, the structural/magnetic phase transitions around 100 K in the pure 10-3-8 compound are suppressed. Bulk superconductivity occurs at 13.3 K at 4.4\% doping, rises to 26.1 K at 14.5\% doping, and then decreases to 22.1 K at 18.2\% doping.

Work at Los Alamos was performed under the auspices of the U.S. Department of Energy, Office of Science, Division of Materials Science and Engineering. Work at Princeton University was supported by the AFOSR MURI on superconductivity. Work at Ames Laboratory (WES, MAT, RP) was supported by the U.S. Department of Energy, Office of Basic Energy Science, Division of Materials Sciences and Engineering. Ames Laboratory is operated for the U.S. Department of Energy by Iowa State University under Contract No. DE-AC02-07CH11358. Dr. Ni acknowledges the Marie Curie Fellowship at Los Alamos National Laboratory. The authors thank Mr. Eunsung Park, Dr. Xin Lu and Dr. Ryan Baumbach for useful discussions.


\begin{thebibliography}{99}
\bibitem{gamma}

We asumme the phonon contribution of the $x=0$ sample and $x=0.145$ sample obey the law of corresponding states. Then the electronic specific heat of the $x=0.145$ sample is obtained using $C_e^{x=0.145}=C_p^{x=0.145}- A\cdot C_{phonon}^{x=0}$, where
$C_{phonon}^{x=0}$ is obtained by subtracting the electronic part from $C_p^{x=0}$. A and $\gamma_n^{x=0.145}$ are thus determined by enforcing $\int _0^{T_C} \gamma_n^{x=0.145} dT=\int _0^{T_C} C_e^{x=0.145}/TdT$.
\end{thebibliography}
\end{document}